%% file: main.tex
\documentclass[11pt,preprint]{aastex}

\usepackage{amsmath}
\usepackage{xspace}
\usepackage{listings}
\usepackage{epsfig}
\usepackage{multirow}
\usepackage{lineno}
\usepackage{color}

\newcommand{\Fermi}[0]{\textit{Fermi}\xspace}
\newcommand{\bbbar}[0]{\ensuremath{b \bar b}\xspace}
\newcommand{\sigmav}[0]{\ensuremath{\langle \sigma v \rangle}\xspace}
\newcommand{\TS}[0]{\ensuremath{\mathrm{TS}}\xspace}

\newcommand{\TSspec}[1]{\ensuremath{\mathrm{TS}_\mathrm{spec}^{#1}}\xspace}
\newcommand{\TSext}[1]{\ensuremath{\mathrm{TS}_\mathrm{ext}^{#1}}\xspace}

\newcommand{\Rmax}[0]{\ensuremath{R_{V_{\rm max}}}\xspace}
\newcommand{\Vmax}[0]{\ensuremath{V_{\rm max}}\xspace}
\newcommand{\Mtidal}[0]{\ensuremath{M_{\rm tidal}}\xspace}
\newcommand{\Msolar}[0]{\ensuremath{M_\odot}\xspace}

\newcommand{\unit}[1]{\ensuremath{\mathrm{\,#1}}\xspace}
\newcommand{\GeV}{\unit{GeV}}
\newcommand{\MeV}{\unit{MeV}}
\newcommand{\degree}{\unit{^{\circ}}}
\newcommand{\cm}{\unit{cm}}
\newcommand{\km}{\unit{km}}
\newcommand{\kpc}{\unit{kpc}}
\newcommand{\second}{\unit{s}}
\newcommand{\photons}{\unit{ph}}
\newcommand{\photon}{\unit{ph}}

\newcommand{\code}[1]{\lstinline!#1!\xspace}
\newcommand{\Sourcelike}[0]{\lstinline!Sourcelike!\xspace}
\newcommand{\gtlike}[0]{\lstinline!gtlike!\xspace}

\newcommand{\gtobssim}[0]{\lstinline!gtobssim!\xspace}

\newcommand{\Figref}[1]{Figure~\ref{fig:#1}}
\newcommand{\Secref}[1]{Section~\ref{sec:#1}}
\newcommand{\Subsecref}[1]{Section~\ref{subsec:#1}}

\newcommand{\Tabref}[1]{Table~\ref{tab:#1}}
\newcommand{\Eqnref}[1]{Equation~(\ref{eqn:#1})}

\newcommand{\lblcaption}[3]{\caption[#3]{{#2}}\label{#1}}
\newcommand{\ApJFigure}[4]{
  \begin{figure}
    \epsscale{#3}
    \plotone{./#1.eps}
    \lblcaption{fig:#4}{#2}{}
  \end{figure}}

\newcommand{\tablestyle}[1]{
  \tabletypesize{\footnotesize}
  \tablecolumns{#1}
  \tablewidth{0pt}
}

\newif\ifcolor
\colortrue

\slugcomment{Accepted by the Astrophysical Journal}

\begin{document}

\title{SEARCH FOR DARK MATTER SATELLITES USING THE FERMI-LAT}
\input{authors.tex}

\newcommand{\highlight}[1]{\colorbox{yellow}{#1}}

\input{abstract.tex}
\maketitle
\input{introduction.tex}
\input{theory.tex}
\input{methods.tex}
\input{results.tex}
\input{interpretation.tex}
\input{conclusion.tex}
\input{acknowledge.tex}
\input{appendix.tex}
\bibliographystyle{apj}
\bibliography{bib}
\input{figures.tex}
\input{tables.tex}

\end{document}

%% file: authors.tex
\author{
M.~Ackermann\altaffilmark{1}, 
A.~Albert\altaffilmark{2}, 
L.~Baldini\altaffilmark{3}, 
J.~Ballet\altaffilmark{4}, 
G.~Barbiellini\altaffilmark{5,6}, 
D.~Bastieri\altaffilmark{7,8}, 
K.~Bechtol\altaffilmark{9}, 
R.~Bellazzini\altaffilmark{3}, 
R.~D.~Blandford\altaffilmark{9}, 
E.~D.~Bloom\altaffilmark{9,10}
% ...
}
\altaffiltext{1}{Deutsches Elektronen Synchrotron DESY, D-15738 Zeuthen, Germany}
\altaffiltext{2}{Department of Physics, Center for Cosmology and Astro-Particle Physics, The Ohio State University, Columbus, OH 43210, USA}
\altaffiltext{3}{Istituto Nazionale di Fisica Nucleare, Sezione di Pisa, I-56127 Pisa, Italy}
\altaffiltext{4}{Laboratoire AIM, CEA-IRFU/CNRS/Universit\'e Paris Diderot, Service d'Astrophysique, CEA Saclay, 91191 Gif sur Yvette, France}
\altaffiltext{5}{Istituto Nazionale di Fisica Nucleare, Sezione di Trieste, I-34127 Trieste, Italy}
\altaffiltext{6}{Dipartimento di Fisica, Universit\`a di Trieste, I-34127 Trieste, Italy}
\altaffiltext{7}{Istituto Nazionale di Fisica Nucleare, Sezione di Padova, I-35131 Padova, Italy}
\altaffiltext{8}{Dipartimento di Fisica ``G. Galilei", Universit\`a di Padova, I-35131 Padova, Italy}
\altaffiltext{9}{W. W. Hansen Experimental Physics Laboratory, Kavli Institute for Particle Astrophysics and Cosmology, Department of Physics and SLAC National Accelerator Laboratory, Stanford University, Stanford, CA 94305, USA}
\altaffiltext{10}{email: elliott@slac.stanford.edu}
% ...

%% file: abstract.tex
\begin{abstract}
Numerical simulations based on the $\Lambda$CDM model of cosmology predict a large number of as yet unobserved Galactic dark matter satellites. We report the results of a Large Area Telescope (LAT) search for these satellites via the $\gamma$-ray emission expected from the annihilation of weakly interacting massive particle (WIMP) dark matter. Some dark matter satellites are expected to have hard $\gamma$-ray spectra, finite angular extents, and a lack of counterparts at other wavelengths. We sought to identify LAT sources with these characteristics, focusing on $\gamma$-ray spectra consistent with WIMP annihilation through the \bbbar channel. We found no viable dark matter satellite candidates using one year of data, and we present a framework for interpreting this result in the context of numerical simulations to constrain the velocity-averaged annihilation cross section for a conventional $100\GeV$ WIMP annihilating through the \bbbar channel.
\end{abstract}


%% file: introduction.tex
\section{Introduction}

Astronomical evidence and theoretical arguments suggest the existence of non-baryonic cold dark matter (CDM).  In standard model cosmology, dark matter constitutes approximately 85\% of the matter density and nearly one-fourth of the total energy density of the universe~\citep{Komatsu:2010fb}. While very little is known about dark matter beyond its gravitational interactions, a popular candidate is a weakly interacting massive particle (WIMP)~\citep{Jungman:1995df,Bergstrom:2000pn,Bertone:2004pz}. From an initial equilibrium state in the hot, dense phase of the early universe, WIMPs freeze-out with a significant relic abundance to constitute much, if not all, of the dark matter in the universe. In regions of high dark matter density, WIMPs may continue to annihilate into Standard Model particles through the same processes that originally set their relic abundance.

Gamma rays produced in the final state of WIMP annihilation, either mono-energetically from direct annihilation or as a continuum of energies through annihilation into intermediate states, may be detectable by the Large Area Telescope (LAT) on board the \textit{Fermi Gamma-ray Space Telescope} (\Fermi).  These $\gamma$ rays would be produced preferentially in regions of high dark matter density within the Milky Way or extragalactic sources. The LAT has reported upper limits on mono-energetic $\gamma$-ray line from annihilations in the smoothly distributed dark matter halo surrounding the Galactic plane~\citep{Abdo:2010nc}. Additionally, upper limits on the continuum $\gamma$-ray emission from  WIMP annihilation have been placed using dwarf spheroidal galaxies~\citep{Abdo:2010ex,Ackermann:2011wa}, the diffuse isotropic background~\citep{Abdo:2010dk}, and clusters of galaxies~\citep{Ackermann:2010rg}. 

Cosmological \textit{N}-body simulations predict that the Milky Way should have many more satellites than are currently observed at optical wavelengths~\citep{Diemand2007,Springel2008}. This prediction allows for the possibility that the majority of these satellites are composed solely of dark matter. While these simulations have the resolution to recover all satellites down to a mass of $\sim 10^6$ \Msolar, the minimum mass of a bound satellite orbiting the Milky Way may be as low as the Earth mass ($\sim 10^{-6}$ \Msolar), a scale that is roughly set by the WIMP velocity at freeze-out~\citep{Hofmann:2001bi,Loeb:2005pm}. Massive dark matter satellites located in the outer regions of the Galactic halo or lower mass satellites located near to Earth may constitute significantly extended $\gamma$-ray sources that could be detectable with the LAT. It has been suggested~\citep{Tasitsiomi:2002vh,Koushiappas:2003bn,Pieri:2007ir,Baltz:2008wd,Springel:2008zz,Anderson:2010df} that a significant $\gamma$-ray flux from WIMP annihilation could arise from these dark matter satellites within the Milky Way halo.

Here, we report on a search for dark matter satellites via $\gamma$-ray emission from WIMP annihilation. We begin by examining the theoretical motivation for such a search. Then, we describe the selection of unassociated, high Galactic latitude $\gamma$-ray sources from both the First LAT Source Catalog (1FGL)~\citep{1FGL} and an independent list of source candidates created with looser assumptions on the source spectrum.  The likelihood ratio test was used to distinguish extended sources from point sources and WIMP annihilation spectra from conventional power-law spectra. No dark matter satellite candidates were found in either the unassociated 1FGL sources or the additional list of candidate sources.  Finally, Via Lactea II~\citep{Diemand2007} and Aquarius~\citep{Springel2008} simulations were used to derive upper limits on the annihilation cross section for a $100\GeV$ WIMP annihilating through the \bbbar channel.


%% file: theory.tex
\section{Dark Matter Satellites}
\label{sec:theory}
In this section, we describe models for the spatial and mass distributions of dark matter satellites in the Galactic dark matter halo, as well as models for the internal density profiles of these satellites. We motivate the search for dark matter satellites lacking optical counterparts by predicting the number of satellites that are detectable, given the constraints on dark matter annihilation from dwarf spheroidal galaxies. We discuss how extrapolating the mass function of dark matter satellites below the mass resolution of current numerical simulations impacts LAT detection potential. 

\subsection{Numerical Simulations}
The Aquarius~\citep{Springel2008} and Via Lactea II (VL-II)~\citep{Diemand2007} projects are currently the highest resolution numerical simulations of dark matter substructure at the scale of Galactic halos. As part of the Aquarius project, six independent simulations of a Milky Way-mass dark matter halo were generated.  The VL-II project provides an additional, independent simulation. These simulations model the formation and evolution of a Milky Way-size dark matter halo and its satellites having over one billion $\sim 10^3$ \Msolar particles. Each simulation resolves over 50,000 satellites within its respective virial radius, which is defined as the radius enclosing an average density 200 times the cosmological mean matter density. Each bound satellite has associated with it a position with respect to the main halo, a velocity, a tidal mass, \Mtidal, a maximum circular velocity, \Vmax, and a radius of maximum circular velocity, \Rmax.

Generally, the $\gamma$-ray flux from annihilating dark matter in a satellite ($\phi_\mathrm{WIMP}$) can be expressed as a product of the line-of-sight integral of the dark matter distribution in a satellite ($J$-factor) and a component depending on the particle physics model ($\Phi^{\rm PP}$) for WIMP annihilation~\citep{Strigari:2006rd}
\begin{equation}
\label{eqn:flux}
\phi_{\mathrm{WIMP}}(E) = J \times \Phi^{\rm PP}(E)
\end{equation}
where
\begin{align}
\label{eqn:jfactor}
J &= \int_0^{\Delta \Omega} \left \{ J(\theta) \right \} d\Omega \nonumber \\
  &= \int_0^{\Delta \Omega}\left \{ \int^{\ }_{\rm l.o.s.}\rho^2[r(D,l,\theta)] dl \right \} d\Omega \\
\label{eqn:japprox}
J &\approx \frac{1}{D^2}\int_V{{\rho (r)}^2dV}
\end{align}
and
\begin{align}
\label{eqn:phipp}
{\Phi }^{\rm PP}(E) &=\frac{1}{4\pi }\frac{\sigmav}{2M^2_{\rm WIMP}} {\sum_f{\frac{dN_f}{dE} B_f} }.
\end{align} 
\Eqnref{jfactor} represents the line-of-sight integral along $l$ through the satellite dark matter density profile $\rho$, where $D$ is the distance to the satellite center, $\theta$ is the offset angle relative to the center, and $r(\theta,D,l) = \sqrt{D^2 + l^2 - 2Dl\cos{\theta}}$ is the distance from the satellite center. The solid angle integral is performed over $\Delta \Omega = 2\pi ( 1 - \cos \theta)$. For satellites at large distances from the Earth, the $J$-factor can be approximated by \Eqnref{japprox} where the volume integration is performed out to the satellite tidal radius~\citep{Tyler:2002ux}. In \Eqnref{phipp}, $\langle \sigma v \rangle$ is the velocity-averaged annihilation cross section, $M_\mathrm{WIMP}$ is the mass of the dark matter particle, and the sum runs over all possible pair annihilation final states with $dN_f/dE$ and $B_f$ representing the photon spectrum and branching ratio, respectively.

For the present analysis, it was important to model the dark matter distribution within the satellites themselves. A Navarro$-$Frenk$-$White (NFW) profile~\citep{Navarro:1996gj} with scale radius $r_s$ and scale density $\rho_s$ was used to approximate the dark matter distribution in the satellites:
\begin{equation}
\rho(r) = \frac{\rho_s}{(r / r_s) (1+r/r_s)^2}. 
\label{eqn:nfw}
\end{equation}
An NFW profile was uniquely defined for each satellite from the values of \Vmax and \Rmax using the relations~\citep{Kuhlen:2008}
\begin{align}
\label{eqn:rs}
r_s      & = \frac{\Rmax}{2.163} \\
\label{eqn:rhos}
\rho_s   & = \frac{4.625}{4\pi G}{\left(\frac{\Vmax}{r_s}\right)}^2
\end{align}  
where $r_s$ is in $\kpc$ and $\rho_s$ is in $\Msolar \kpc^{-3}$.  The existence of dark matter substructure within the satellites themselves would increase the $J$-factor, but this contribution is expected to be no greater than a factor of a few~\citep{Martinez:2009jh,Anderson:2010df}.  Thus, we took a conservative approach and did not include this enhancement when calculating $J$-factors.

Although the NFW model is widely used for its simplicity and is broadly consistent with the dark matter distribution in satellites~\citep{Springel2008}, a few caveats are important. First, simulations with increased resolution have revealed that more scatter exists in the central dark matter density profile than is implied by the NFW profile~\citep{Navarro:2008kc}. In fact, these simulations indicate that the central dark matter density in satellites is systematically shallower than the $r^{-1}$ central density implied by the NFW model. However, these deviations occur at a scale of $\lesssim 10^{-3}$ of the halo virial radius~\citep{Navarro:2008kc}, and thus do not strongly affect the predicted flux given the LAT's angular resolution. Second, satellites are, in nearly all cases, more severely tidally truncated than the $r^{-3}$ outer density scaling of the NFW profile. For satellites with large $r_s$, using the NFW profile will result in a slight ($<10\%$) overestimation of the predicted flux.

\subsection{Extrapolation to Low-mass Satellites}
\label{lowmass}
The satellite mass functions for the Aquarius and VL-II simulations are complete down to $\sim 10^6$ \Msolar. However, theoretical arguments suggest that the mass function of satellites may extend well beyond this resolution limit, perhaps down to Earth-mass scale ($\sim 10^{-6}$ \Msolar). Therefore, it is likely that the satellites resolved in the Aquarius and VL-II simulations are only a small fraction of the total number of bound dark matter satellites present in our Galaxy. We refer to these unresolved dark matter satellites with $\Mtidal < 10^6\Msolar$ as ``low-mass'' satellites.

To estimate the contribution of low-mass satellites to the LAT search, we extrapolated the distribution of satellites in VL-II down to $1 \Msolar$. Assuming a power-law mass function for satellites, $dN/d\Mtidal \propto \Mtidal^{-1.90}$~\citep{Madau2008,Springel:2008zz}, we calculated the number of satellites at a given \Mtidal within 50 kpc of the Galactic center. These low-mass satellites were distributed within this 50 kpc radius in accordance to the radial distribution described in Madau et al. (2008). The cut at 50 kpc is conservative based on the null-detection of the Segue 1 dwarf spheroidal with a tidal mass of $\sim 10^7 \Msolar$ at $\sim 28 \kpc$ from the Galactic center~\citep{Geha:2008zr}.

In order to model the internal dark matter distribution of low-mass satellites with an NFW profile (using Equation~(\ref{eqn:rs}) and (\ref{eqn:rhos})), we fit relationships between \Mtidal, \Vmax, and \Rmax. For VL-II satellites within 50 kpc of the Galactic center and with $\Mtidal > 10^6\Msolar$, we found that 
\begin{equation}
\label{eqn:vmax}
\Vmax = V_0 \left ( \frac{\Mtidal}{\Msolar} \right)^{\beta}
\end{equation}
with $V_0 = 10^{-1.20 \pm 0.05} \km \second^{-1}$, $\beta = 0.30 \pm 0.01$, and a log-Gaussian scatter of $\sigma_{\Vmax} = 0.063 \km \second^{-1}$. Additionally, we found that 
\begin{equation}
\label{eqn:rvmax}
\Rmax = R_0 \left( \frac{\Mtidal}{\Msolar} \right)^{\delta}
\end{equation}
with $R_0 = 10^{-3.1 \pm 0.4} \kpc$, $\delta=0.39 \pm 0.02$, and a log-Gaussian scatter of $\sigma_{\Rmax} = 0.136 \kpc$. Using these relationships, we randomly generated low-mass satellites consistent with the VL-II simulation down to a tidal mass of $1\Msolar$.

\subsection{Comparison to Optical Satellites} 
\label{subsec:lowmass}
The VL-II simulation, extrapolated as described above, provides a theoretical model for the population of Milky Way dark matter satellites from $10^{10}\Msolar$ to $1\Msolar$. A simple estimate of the detectable fraction of these dark matter satellites can be obtained from the 11-month limits on the WIMP annihilation flux from dwarf spheroidals~\citep{Abdo:2010ex}. No $\gamma$-ray signal was detected and the strongest limits on the annihilation cross section result from the analysis of the Draco dwarf spheroidal, which has a $J$-factor (integrated over the solid angle of a cone with radius $0.5\degree$) of $\sim 10^{19}\GeV^2\cm^{-5}$. The central mass of Draco is well known from stellar kinematics~\citep{Strigari:2008ib,Walker:2009zp,Wolf:2009tu}; however, the total dark matter mass for Draco is less certain due to the lack of kinematic measurements in the outer regions of the halo. A conservative lower bound on the total dark matter halo mass of Draco was taken to be $10^8\Msolar$. Based on this lower bound, we determined what fraction of the satellites have a larger detection potential than Draco, i.e., what fraction of satellites have a $J$-factor greater than that of Draco.

For an NFW profile, the $J$-factor is roughly described by
\begin{equation}
\label{eqn:massext}
J \propto \frac{r^3_s{\rho }^2_s}{D^2} \propto \frac{M^{0.81}}{D^2}.
\end{equation}
using the relationships that $r_s\propto M^{0.39}$ (from Equations~(\ref{eqn:rs}) and (\ref{eqn:rvmax})) and ${\rho }_s\propto M^{-0.18}$ (from Equations~(\ref{eqn:rs})$-$(\ref{eqn:vmax})). \Eqnref{massext} makes it possible to compare the relative astrophysical contribution to the $\gamma$-ray flux for different halos based on their tidal mass and distance from Earth (\Figref{image13}). The choice of particle physics annihilation model merely scales all satellites by the same constant factor.

\Figref{image13} serves as a guide for evaluating the detectability of low-mass satellites. While the total number of satellites increases with decreasing mass, the $J$-factors of these low-mass satellites tend to decrease. This means that low-mass satellites, while dominating the local volume in number, are a subdominant contributor to the $\gamma$-ray flux at the Earth. Using the procedure discussed in \Secref{interpretation}, we verify that extending the VL-II mass function to low mass has a minimal effect ($<5\%$) when setting upper limits on \sigmav; consequently, we do not consider these satellites in our primary analysis. For low-mass satellites to dominate the $\gamma$-ray signal, a mechanism must be invoked to either increase the concentration for low-mass satellites, or decrease the slope of the mass function. Of course, the above statements do not preclude the possibility that there could be a low-mass satellite with a high $J$-factor very near to the Sun. 

In the context of the CDM theory, several dark satellite galaxies with no associated optical emission could be detectable by the LAT. In addition to motivating our satellite search, \Figref{image13} allowed us to narrow our focus to those satellites with the best prospects for detection. Using \Eqnref{massext}, we omitted satellites with $J$-factors more than an order of magnitude less than the lower bound on the $J$-factor for Draco. This greatly reduced the number of satellites for which the full line-of-sight integral in \Eqnref{jfactor} was calculated.


%% file: methods.tex
\section{Methods} 
\label{sec:methods}
In this section, we review the tests applied to LAT sources lacking associations in other wavelengths to determine if any are consistent with dark matter satellites.  First, we summarize our data set and give an overview of an independent search for LAT sources without spectral assumptions.  Then, we define a procedure for selecting candidate dark matter satellites using the likelihood ratio test to evaluate the spatial extension and spectral shape of each source.  The ability of the LAT to detect spatial extension and spectral shape depends on source flux and spectral hardness. Extensive Monte Carlo simulations were required to determine cuts for rejecting both point-like sources and sources with power-law spectra at 99\% confidence.  When combined, these cuts allowed us to select for extended, non-power-law sources in our sample of high-latitude unassociated LAT source candidates with a contamination level of 1 in $10^{4}$. The work presented in this section is largely derived from Wang (2011).

\subsection{Data Selection}
\label{subsec:data}
\Fermi has been operating in sky-scanning survey mode since early 2008 August. The primary instrument on board \Fermi is the LAT, designed to be sensitive to $\gamma$-rays in the range from $20\MeV$ to $>\,300\GeV$. The LAT has unprecedented angular resolution and sensitivity in this energy range, making it an excellent instrument for detecting new $\gamma$-ray sources~\citep{Atwood:2009ez}.

Our data sample consisted of ``Diffuse'' class events from the first year of LAT data collection (2008 August 8 to 2009 August 7), which overlapped substantially with the data used for the 1FGL~\citep{1FGL}.  To reduce $\gamma$-ray contamination from the bright limb of the Earth, we rejected events with zenith angles larger than 105\degree and events taken during time periods when the rocking angle of the LAT was greater than 47\degree (the nominal LAT rocking angle was $35$\degree during this time period).  Due to calibration uncertainties at low energy and the current statistical limitations in the study of the instrument response functions (IRFs) above $300\GeV$, we accepted only photons with energies between $200\MeV$ and $300\GeV$.  This analysis was limited to sources with Galactic latitudes greater than 20\degree, since the Galactic diffuse emission complicates source detection and the analysis of spatial extension at lower Galactic latitudes.  We modeled the diffuse $\gamma$-ray emission with standard Galactic (\textit{gll\_iem\_v02.fit}) and isotropic (\textit{isotropic\_iem\_v02.txt}) background models\footnote{ http://fermi.gsfc.nasa.gov/ssc/data/access/lat/BackgroundModels.html}.  Throughout this analysis, we used the LAT ScienceTools\footnote{http://fermi.gsfc.nasa.gov/ssc/data/analysis/software} version v9r18p1 and the P6\_V3\_DIFFUSE IRFs\footnote{http://fermi.gsfc.nasa.gov/ssc/data/analysis/scitools/overview.html}.

\subsection{Source Selection}
\label{sec:search}
The 1FGL is a collection of high-energy $\gamma$-ray sources detected by the LAT during the first 11 months of data taking~\citep{1FGL}.  It contains 1451 sources, of which 806 are at high Galactic latitude ($|b|>20\degree$).  Of these high-latitude 1FGL sources, 231 are unassociated with sources at other wavelengths and constitute the majority of the sources tested for consistency with the dark matter satellite hypothesis.  However, the 1FGL spectral analysis, including the threshold for source acceptance, assumed that sources were point-like with power-law spectra. This decreased the sensitivity of the 1FGL to both spatially extended and non-power-law sources, which are characteristics expected for dark matter satellites.  In an attempt to mitigate these biases, we augmented the unassociated sources in the 1FGL with an independent search of the high-latitude sky.

We performed a search for $\gamma$-ray sources using the internal LAT Collaboration software package, \Sourcelike~\citep{Abdo:2010qd,Grondin:2011kw}.  \Sourcelike performs a fully binned likelihood fit in two dimensions of space and one dimension of energy.  When fitting the spectrum of a source, \Sourcelike fits the fraction of counts associated to the source in each energy bin independently. The overall likelihood is the product of the likelihoods in each bin.  This likelihood calculation has more degrees of freedom than that performed by the LAT ScienceTool, \gtlike, which calculates the likelihood from all energy bins simultaneously according to a user-supplied spectral model. In this analysis, we used 11 energy bins logarithmically spaced from $200\MeV$ to $300\GeV$.

Using \Sourcelike, we searched for sources in 2496 regions of interest (ROIs) of dimension $10\degree \times 10\degree$ centered on HEALPix~\citep{Gorski:2005} pixels obtained from an order four tessellation of the high-latitude sky ($|b| > 20\degree$).  Each  ROI was sub-divided into $0.1\degree \times 0.1\degree$ pixels, and for each pixel the likelihood of a point source at that location was evaluated by comparing the maximum likelihood ($\mathcal{L}$) of two hypotheses: (1) that the data were described by the standard LAT diffuse background models without any point sources ($H_0$), and (2) that the data were described by the existing model with an additional free parameter corresponding to the flux of a source at the target location ($H_1$).  Utilizing the likelihood ratio test, we defined a test statistic (TS):
\begin{equation}
  \TS = - 2 \ln \left ( \frac{\mathcal{L}(H_0)}{\mathcal{L}(H_1)} \right ). 
\end{equation}

After generating a map of the test statistic over the entire high-latitude sky, we iteratively refit regions around potential source candidates more carefully. The flux normalizations, spectral indices, and emission centroids of candidate sources with $\TS > 16$ were refined while incorporating the flux normalizations of diffuse backgrounds and other candidate point sources with $\TS>16$ within the ROI as free parameters in the fit.  After refitting, only candidate sources with $\TS > 24$ were accepted into the list of source candidates\footnote{Monte Carlo simulations have shown that 1 in $10^4$ background fluctuations will be detected at $\TS \geq 24$ when fit with \Sourcelike.~\citep{Wang:2011}.}.  Finally, to avoid duplicating sources in the 1FGL, we removed candidate sources with 68\% localization errors overlapping the 95\% error ellipse given for 1FGL sources.

Our search of the high-latitude sky revealed 710 candidate sources, of which 154 were not in the 1FGL (36 of these candidate sources subsequently appear in the Second LAT Source Catalog~\citep{2FGL}).  We did not expect to recover all 806 high-latitude 1FGL sources, since the 1FGL is a union of four different detection methods and external seeds from the BZCAT and WMAP catalogs~\citep{1FGL}.  However, since \Sourcelike fits each energy bin independently, we expected to find source candidates that were excluded from the 1FGL, either because they had non-power-law spectra or they had hard spectra with too few photons to pass the 1FGL spectral analysis.  We sacrificed some sample purity for detection efficiency in our candidate source list because stringent cuts on spatial extent and spectral shape were later applied.  We obtained a final list of 385 high-latitude unassociated LAT sources and source candidates by combining the 231 unassociated sources in the 1FGL with these 154 non-1FGL candidate sources.

To check for consistency with the source analysis of the 1FGL, we performed an unbinned likelihood analysis with \gtlike assuming that the unassociated sources were point-like with power-law spectra.  Our fitted fluxes and spectral indices are in good agreement with those in the 1FGL for the 231 unassociated 1FGL sources. The values are plotted in \Figref{image1}, where it can be seen that the unassociated LAT sources span a wide range of fluxes and spectral indices.  This wide range was taken into account when designing selection criteria for candidate dark matter satellites.   The strong correlation between spectral index and flux is due to the improvement of the point-spread function (PSF) of the LAT with increasing energy and the relatively soft spectral dependence of the Galactic diffuse background.  It is apparent that there are more non-1FGL source candidates in this sample with very hard spectra (spectral index $ \sim 1.0$) and very low fluxes ($\sim 10^{-10} \photons \cm^{-2} \second^{-1}$). In the Appendix, we show that these source candidates are very likely spurious.

\subsection{Spatial Extension Test}
\label{subsec:exttest}
The LAT has the potential to resolve some dark matter satellites as spatially extended $\gamma$-ray sources. While the bulk of satellites are not spatially resolvable by the LAT, spatial extension is an important feature for distinguishing large or nearby satellites from point-like astrophysical sources (see \Secref{theory}).  Assuming that the spatial and spectral distribution of $\gamma$ rays produced from dark matter annihilation factorize, the shape of the projected dark matter distribution (\Eqnref{jfactor}) is convolved with the LAT PSF. For an NFW dark matter distribution (\Eqnref{nfw}) with scale radius $r_s$ at a distance $D$, the angular extent of a satellite can be characterized by the parameter $\alpha_0 = r_s/D$. Approximately $90\%$ of the integrated $J$-factor comes from within the angular radius $\alpha_0$~\citep{Strigari:2006rd} and the LAT is sensitive to the spatial extension of satellites with $\alpha > 0.5\degree$.

We used the likelihood ratio test, as implemented by \Sourcelike, to test sources for spatial extension. We defined a test statistic for extension as
\begin{align}
\TSext{} &= - 2 \ln \left ( \frac{\mathcal{L}(H_\mathrm{point})}{\mathcal{L}(H_\mathrm{NFW})} \right ) \\
&= \TS_\mathrm{NFW}-\TS_\mathrm{point}
\end{align}
where $\TS_\mathrm{point}$ was the test statistic of the candidate source assuming that it had negligible extension ($\alpha_0$ much smaller than the LAT PSF) and $\TS_\mathrm{NFW}$ was the test statistic of the candidate source when $\alpha_0$ was fit as a free parameter.  In both cases, the position of the source was optimized during the fit.

We sought to define a cut on the value of \TSext{} to eliminate 99\% of point sources over the range of spectral indices and fluxes found in the unassociated LAT sources.  This cut was labeled \TSext{99}. While the point and extended hypotheses are nested and \TSext{} is cast as a likelihood ratio test, it is unclear whether this analysis satisfied all of the suitable conditions for the application the theorems of Wilks~\citep{Wilks:1938} or Chernoff~\citep{Chernoff:1954}.  Therefore, we relied on simulations to parameterize \TSext{99} as a function of source flux and spectral index.

To evaluate \TSext{99} over the pertinent range of source fluxes and spectral indices, the unassociated LAT sources were bracketed with 10 representative power-law models (\Tabref{tests} and blue stars in \Figref{image1}).  For each of the 10 representative models, 1000 independent sources were simulated at random locations in the high-latitude sky using the LAT simulation tool, \gtobssim, and the spacecraft pointing history for our one-year data set.  To accurately incorporate imperfect modeling of background point and diffuse sources, we embedded the simulated point sources in one year of LAT data and calculated \TSext{} for each simulated source.  We defined \TSext{99} for each representative model as the smallest value of \TSext{} that was larger than that calculated for 99\% of simulated point sources.  The values of \TSext{99} for all 10 models were calculated independently of the $\TS > 24$ detection cut (\Tabref{tests}).

We used a bilinear interpolation to estimate the value of \TSext{99} for any point in the space spanned by the grid of flux and spectral index. Since each measurement of source flux and spectral index has a statistical uncertainty, we interpolated to the largest value of \TSext{99} that was consistent with the $\pm 1\sigma$ error for each source for a conservative estimate of \TSext{99}. 

\subsection{Spectral Test}
\label{subsec:spectest}
To select sources that were spectrally consistent with WIMP dark matter annihilation, we designed a test for spectral curvature.  The continuum $\gamma$-ray emission from WIMP annihilation has two different contributions: secondary photons from tree-level annihilation~\citep{Baltz:2008wd,Cesarini:2003nr} and additional photons from QED corrections -- i.e., final state radiation (FSR)~\citep{Beacom:2004pe}.  For tree-level annihilations, the leading channels among the kinematically allowed final states are predicted to be \bbbar, $t\overline{t}$, $W^+W^-$, $Z^0Z^0$, and ${\tau }^+{\tau }^-$.  The $\gamma$-ray spectra from these channels are quite similar, except for the $\tau $-channel which is considerably harder~\citep{Cesarini:2003nr}. We chose the \bbbar channel as a representative proxy for the tree-level annihilation spectrum.  

We defined a test statistic to evaluate the consistency of the data with dark matter and power-law spectra. This spectral test statistic,
\begin{align}
\TSspec{} &= - 2 \ln \left ( \frac{\mathcal{L}(H_\mathrm{pwl})}{\mathcal{L}(H_{\bbbar})} \right ) \nonumber \\
          &= \TS_{\bbbar} - \TS_\mathrm{pwl}
\end{align}
was the difference in source \TS calculated with an unbinned analysis using \gtlike assuming a \bbbar dark matter spectral model ($\TS_{\bbbar}$) and a power-law ($\TS_\mathrm{pwl}$) spectral model.  These two hypotheses are not nested, and thus the significance of this test was evaluated with simulations.  When performing our fits, we modeled candidate sources as point-like\footnote{From simulations, this has been found to be a conservative way to estimate $\TSspec{}$~\citep{Wang:2011}.} and left their fluxes and the fluxes of the diffuse backgrounds free. Additionally, the power-law and dark matter spectral models contain a spectral free parameter (the dark matter mass or power-law index).

Using the same representative simulations described in \Subsecref{exttest}, we defined \TSspec{99} to be the value of \TSspec{} which was larger than that calculated for 99\% of simulated power-law sources (\Tabref{tests}).  When calculating \TSspec{99} for a particular source, we chose the largest value from a bilinear interpolation to the $\pm 1 \sigma$ errors on fitted flux and spectral index (as discussed at the end of \Subsecref{exttest}). These tests of spatial extension and spectral character allowed us to select non-point-like and non-power-law sources with a contamination of 1 in $10^{4}$ assuming they were independent.


%% file: results.tex
\section{Results}
\label{sec:results}

\subsection{Search for Dark Matter Satellites}
\label{subsec:extsources}

We applied the cuts on spatial extension and spectral character to select DM satellite candidates from the 385 unassociated high-latitude LAT sources and source candidates. Two of the 385 unassociated sources, 1FGL\,J1302.3$-$3255 and 1FGL\,J2325.8$-$4043, passed the cut on spatial extension.  One of these, 1FGL\,J1302.3$-$3255, also passed our spectral test, preferring a \bbbar spectrum to a power-law spectrum.  However, we do not believe that either of these sources is a viable DM satellite candidate for reasons discussed below.

As their names imply, both 1FGL\,J1302.3$-$3255 and 1FGL\,J2325.8$-$4043 are present in the 1FGL~\citep{1FGL}; we summarize information about each source in \Tabref{extsources}.  While 1FGL\,J1302.3$-$3255 was unassociated when the 1FGL was published, and has previously been considered as a promising dark matter satellite candidate~\citep{Buckley:2010}, it has since been associated with a millisecond pulsar by radio follow-up observation~\citep{Hessels:2011vc}. The other possibly extended source, 1FGL\,J2325.8$-$4043, has a high probability of association with two AGN in the first LAT AGN Catalog~\citep{Abdo:2010ge}. 1FGL\,J2325.8$-$4043 is assigned a 70\% probability of association to 1ES\,2322$-$409  and a 55\% probability of association with PKS\,2322$-$411. Additionally, each of these sources was checked for consistency with an $E^{-1}$ power-law spectrum, as would be expected from some FSR models~\citep{Essig:2009jx}, and both can be excluded with 99\% confidence~\citep{Wang:2011}. We further discuss sources with hard spectral indices in the Appendix.

Since AGN are not expected to be spatially extended at an angular scale resolvable by the LAT, we cross checked 1FGL\,J2325.8$-$4043 against the Second LAT Source Catalog~\citep{2FGL}.  In two years of data, two sources were found within 0.5\degree of the location of 1FGL\,J2325.8$-$4043. In one year of data, these two sources could not be spatially resolved, but their existence was enough to favor an extended source hypothesis.  Spurious measurements of a finite extent for point sources are not unexpected. Testing extension with a purity of 99\%, the Poisson probability of finding at least one spurious source in our 385 tests is 98\%. Since 1FGL\,J1302.3$-$3255 is associated with a pulsar and 1FGL\,J2325.8$-$4043 does not appear to be truly extended, we conclude that there were no unassociated, high-latitude spatially extended $\gamma$-ray sources in the first year of LAT data. Thus, according to the criteria defined in \Secref{methods}, no viable dark matter satellite candidates were found.

\subsection{Contamination from Pulsars}

To better understand possible misidentification of pulsars (such as 1FGL\,J1302.3$-$3255) as candidate dark matter satellites, we applied our test of spectral shape to 25 high-latitude LAT detected pulsars.   Of these 25 pulsars, 14 were identified when the 1FGL was published and 11 were subsequently identified through follow-up observations by radio telescope~\citep{Hessels:2011vc}. 

Interestingly, 24 of these pulsars passed our spectral cut, preferring a \bbbar spectrum to a power-law spectrum.  This can be understood by comparing the exponentially cutoff power-law model commonly used to fit pulsars with a \bbbar WIMP annihilation spectrum.  The exponentially cutoff power law has the form~\citep{Abdo:2009ax}:
\begin{equation}
\label{expcutoff}
\frac{dN}{dE}=KE^{-\Gamma }_{\GeV}{\exp  (-\frac{E}{E_\mathrm{cut}})\ } 
\end{equation} 
where $\Gamma$ is the photon index at low energy, $E_\mathrm{cut}$ is the cutoff energy, and $K$ is a normalization factor (in units of $\photon \cm^{-2} \second^{-1} \MeV^{-1}$).  In \Figref{image7}, we plot both the exponentially cutoff power-law model and a low-mass ($M_\mathrm{WIMP} \sim 25\GeV$) \bbbar spectrum and show that for $E > 200\MeV$ the two curves are very similar.  

By fitting \bbbar spectra to the 25 LAT-detected pulsars, we found that they tend to be best fit by low dark matter masses (\Figref{image8}). Although our statistics are limited, the distribution peaks around $30\GeV$, with most pulsars having a best-fit dark matter mass $M_\mathrm{WIMP}<60\GeV$.  This suggests that unidentified, high-latitude pulsars can present a source of confusion in spectral searches for dark matter satellites.  In general, many unassociated LAT sources have spectra that are inconsistent with a power-law model~\citep{Bonamente:2010,2FGL}.  The fact that these sources passed our spectral test does not imply that they are best fit by \bbbar spectra, merely that \bbbar spectra fit better than a simple power law. These unassociated, non-power-law sources were not found to share a consistent spectrum, as would be expected from dark matter annihilation.

The abundance of non-power-law $\gamma$-ray sources emphasizes the importance of testing for spatial extension when attempting to identify dark matter satellites at high latitudes. Some concerns remain due to the fact that the LAT detects spatially extended pulsar wind nebulae located around some pulsars~\citep{Ackermann:2010eu}.  However, we do not expect the older pulsars at high Galactic latitudes to have nebulae that are spatially extended on a scale detectable by the LAT.  Of course, there is always a risk that a chance coincidence with a low-flux neighboring source will cause apparent source extension.


%% file: interpretation.tex
\section{Interpretation in the Context of $N$-body Simulations}
\label{sec:interpretation}
No high-latitude unassociated LAT source candidates passed our dark matter satellite selection criteria. This is combined with the simulations in \Secref{theory} to constrain a conventional $100\GeV$ WIMP annihilating through the \bbbar channel. Monte Carlo simulations were used to determine the detection efficiency for dark matter satellites as a function of flux and spatial extension. For multiple realizations of each $N$-body simulation, we calculated the probability of detecting no satellites from the detection efficiency of each simulated satellite. Averaging over these simulations and increasing \sigmav until the probability of detecting no satellites drops below 5\%, we were able to set a $95\%$ confidence upper limit on \sigmav.

\subsection{Detection Efficiency}
\label{subsec:efficiency}
The detection efficiency of our selection was defined as the fraction of dark matter satellites that pass the cuts in \Secref{methods} and was calculated from Monte Carlo simulations.  The efficiency for detecting a dark matter satellite depended on spectral shape (i.e., dark matter mass and annihilation channel), flux, and spatial extension.  For a $100\GeV$ WIMP annihilating through the \bbbar channel, we examined the efficiency for satellites with characteristic fluxes ranging from $5.0 \times 10^{-11}\photons \cm^{-2} \second^{-1}$ to $5.0 \times 10^{-8} \photons \cm^{-2} \second^{-1}$ and characteristic spatial extension (as described in \Subsecref{exttest}) from $0.5\degree$ to $2.0\degree$.  These ranges were chosen to reflect the fluxes of the unassociated high-latitude LAT sources and angular extents to which the LAT is sensitive\footnote{The 68\% containment radius of the LAT PSF, which depends on photon energy and angle of incidence, can be approximated by the function, $0.8\degree {(\frac{E}{1\GeV})}^{-0.8}$~\citep{1FGL}, yielding $\sim 0.8\degree$ at $1\GeV$ and $\sim 0.13\degree$ at $10\GeV$.}.

For each set of characteristics listed in \Tabref{bbbar}, we utilized \gtobssim to simulate 200 dark matter with NFW profiles and \bbbar spectra from a $100\GeV$ WIMP.  These simulations were embedded in LAT data at random high-latitude locations, and \Sourcelike was used to compute \TSext{}, \TSspec{}, and the detection \TS for each.  The satellite detection efficiency was computed as the fraction of satellites with \Sourcelike $\TS > 24$, $\TSext{} > \TSext{99}$, and $\TSspec{} > \TSspec{99}$.  The first requirement was included as a proxy for the efficiency of the source finding algorithm. The creation of this efficiency table (\Tabref{bbbar}) was computationally intensive and the result is likely model dependent, which limited this analysis to the examination of only the $100\GeV$ \bbbar model. To expedite the generation of this table, we found the flux value with efficiency $<0.05$ and conservatively set the efficiency for sources with less flux to 0.
\subsection{Simulated Satellite Distributions}
\label{subsec:distribution}
The VL-II and Aquarius simulations (described in \Secref{theory}) were used to predict the Galactic dark matter satellite population.  Picking a vantage point 8.5 kpc from the center of each simulation (the solar radius), we calculated the spatial extension and integrated $J$-factor for each satellite.  To account for variation in the local satellite population, we repeated this procedure, creating six realizations from maximally separated vantage points at the solar radius.  It is important to note that while the VL-II and six Aquarius simulations are statistically independent, these different realizations are not -- i.e., the same satellites appear in multiple realizations. Thus, we have seven independent simulations, each with six not-independent realizations (collectively referred to as 42 ``visualizations'').

After excluding undetectably faint satellites with $J$-factors an order of magnitude less than the lower bound the $J$-factor of Draco (\Subsecref{lowmass}), we compared the distributions of $J$-factors and spatial extensions across the visualizations.  We found reasonable agreement in the detectable satellite distributions between the VL-II and Aquarius simulations. The variation in satellite number in bins of flux and spatial extension was much larger between different simulations than between realizations of the same simulation (as is expected, because the realizations are not independent). On average, satellites with $\alpha_0 > 0.5\degree$ make up $\sim30\%$ of the total integrated $J$-factor from satellites in these simulations.

\subsection{Upper Limits}
\label{subsec:upperlimits}

For each of the 42 visualizations of VL-II and Aquarius, we calculated the $\gamma$-ray fluxes of all satellites for a given \sigmav using \Eqnref{flux}. With these fluxes and the true spatial extension for each satellite, we performed a bilinear interpolation on \Tabref{bbbar} to determine the detection efficiency for each satellite. The probability that the LAT would observe none of the satellites in visualization $i$ is
\begin{equation}
\label{eqn:prob}
P_i( \sigmav ) = \prod_{j} (1 - \epsilon_{i,j}( \sigmav ))
\end{equation}
where $\epsilon_{i,j}$ is the detection efficiency for satellite $j$ in visualization $i$. Because there is no reason to favor any one visualization, we calculated the average null detection probability over the $N=42$ visualizations as
\begin{equation}
\bar P (\sigmav) = \frac{1}{N} \sum_i^N P_i( \sigmav )
\end{equation}
To set an upper limit on the dark matter annihilation cross section, we increased \sigmav until the probability of a null observation was $<5\%$, i.e. $\bar P < 0.05$. This corresponds to 95\% probability that, for this \sigmav, at least one satellite would have passed our selection criteria. Using this methodology, the LAT null detection constrains \sigmav to be less than $1.95 \times 10^{-24} \cm^{3} \second^{-1}$ for a $100\GeV$ WIMP annihilating through the \bbbar channel.


%% file: conclusion.tex
\section{Discussion and Conclusions}

We performed a search for dark matter satellites using one year of \Fermi-LAT data. After completing an independent search for $\gamma$-ray sources, we constructed tests to evaluate both source extension and spectral shape. We distinguished dark matter satellite candidates by selecting spatially extended sources with $\gamma$-ray spectra consistent with those produced by dark matter particle annihilation to \bbbar.  Our initial scans selected two potentially extended sources; however, follow-up analyses revealed that neither of them is a valid dark matter satellite candidate. Therefore, we concluded that, given our pre-defined search criteria, there were no signals of dark matter satellites in the first year LAT data. 

Using $\Lambda$CDM-based theoretical predictions from the Aquarius and Via Lactea II numerical simulations of the Galactic dark matter distribution, we estimated the number of dark matter satellites that could be observed assuming a $100\GeV$ \bbbar model with varying cross section. We quantified the detection efficiency for these satellites, and used it to set an upper limit on the velocity averaged annihilation cross section of $1.95 \times 10^{-24} \cm^{3} \second^{-1}$ for a $100\GeV$ WIMP annihilating to $\bbbar$. This limit is approximately 60 times greater than the expected value of the thermal relic cross section, and it is about an order of magnitude less stringent than the best one-year limit from known dwarf spheroidals~\citep{Abdo:2010ex}. This difference in upper limits can be accounted for by the fact that the analysis of dwarf spheroidals effectively probes the flux limit of the LAT detector, while our selection on spatial extension and spectral shape limits us to higher source fluxes.

We have presented a novel technique to search for dark matter satellites using the LAT. Further data will improve the prospects for finding dark matter satellites using the methodology developed in this paper. Higher quality data and analysis techniques are likely to reveal the presence of more unassociated sources. Improved analysis techniques will make it possible to select dark matter candidates with increased efficiency while maintaining the same discrimination power. Additionally, using the analysis in this paper as a guide, one can consider a broader class of dark matter annihilation models.


%% file: acknowledge.tex
\section{Acknowledgements}

The \textit{Fermi} LAT Collaboration acknowledges generous ongoing support from a number of agencies and institutes that have supported both the development and the operation of the LAT as well as scientific data analysis. These include the National Aeronautics and Space Administration and the Department of Energy in the United States, the Commissariat \`a l'Energie Atomique and the Centre National de la Recherche Scientifique / Institut National de Physique Nucl\'eaire et de Physique des Particules in France, the Agenzia Spaziale Italiana and the Istituto Nazionale di Fisica Nucleare in Italy, the Ministry of Education, Culture, Sports, Science and Technology (MEXT), High Energy Accelerator Research Organization (KEK) and Japan Aerospace Exploration Agency (JAXA) in Japan, and the K.~A.~Wallenberg Foundation, the Swedish Research Council and the Swedish National Space Board in Sweden. Support was also provided by the Department of Energy Office of Science Graduate Fellowship Program (DOE SCGF) administered by ORISE-ORAU under contract no. DE-AC05-06OR23100.

Additional support for science analysis during the operations phase is gratefully acknowledged from the Istituto Nazionale di Astrofisica in Italy and the Centre National d'\'Etudes Spatiales in France.

%% file: appendix.tex
\appendix

\section{Investigating High Latitude Spurious Sources}
\label{appsec:spurious}
In this Appendix, we estimate the number and distribution of spurious sources arising from our source search method. The TS of each source can be related to the probability that such an excess can be obtained from background fluctuations alone. It was shown~\citep{Mattox:1996} that, for the EGRET $\gamma$-ray telescope, the distribution of TS at a fixed point for background-only simulated data followed a $\chi^2/2$ distribution with one degree of freedom as expected from Chernoff's theorem~\citep{Chernoff:1954}. Once convinced that the necessary conditions were met for Chernoff's theorem, it was possible to use standard combinatorial arguments, given the method used to scan the sky, to calculate the false source rate at any TS value. However, in our analysis it was unclear whether all of the suitable conditions of Chernoff's theorem were satisfied. Indeed, Wang (2011) shows that, in our case, the source TS distribution does not follow that predicted by Chernoff's theorem.

To estimate the number of spurious sources in our candidate source list, we applied our source search procedure to a one-year background-only Monte Carlo simulation (Galactic and isotropic diffuse only).  The one-year LAT observation simulation was generated using \gtobssim with the actual LAT pointing history and application of the same event selection (P6\_V3\_DIFFUSE IRFs, time range, energy range, zenith angle and rocking angle).

We searched the background-only simulated data at $|b|>20\degree$ using \Sourcelike, and found 193 candidate sources with $\TS > 24$. Of course, in this case all 193 sources were spurious. This distribution was uniform over the high-latitude sky and was consistent with arising from random fluctuations in the background. The locations of the spurious sources have no coincidences with any of the 385 unassociated sources and source candidates found in the LAT data (see \Secref{search}).

The 193 spurious sources were individually analyzed with \gtlike, using a point-source spatial model and a power-law spectral model. \Figref{image3} shows the distribution of the spectral index and the integral flux from $200\MeV$ to $300\GeV$ for these spurious sources (blue crosses) and the 385 unassociated sources and source candidates (black and red squares). The same correlation between spectral index and flux holds for both the Monte Carlo simulation and LAT data.

We specifically studied the spurious sources and unassociated sources with spectral indices in the range from 1.0 to 0.7 since a cluster of sources in this range are predicted by some leptophilic dark matter models~\citep{Arvanitaki:2009, Grasso:2009}. In this range of spectral indices, there are 12 spurious sources in the Monte Carlo simulation and 13 unassociated source candidates in the LAT data. We applied a more stringent set of photon selection cuts, to obtain a cleaner photon sample than the ``Diffuse'' class~\citep{Abdo:2010nc}. This cut removed most of the residual high-energy charged-particle background from the regions of the 13 unassociated source candidates and resulted in a general reduction of source significance and the disappearance of two sources ($\TS \sim 0$).  This implied that the detection of some of these unassociated hard source candidates was due to high-energy charged particle contamination. There were no obvious differences in the spatial or spectral distribution of stringently selected photons in the regions of the 13 unassociated source candidates and simulated photons from the 12 spurious sources.  Additionally, we reanalyze these 13 unassociated source candidates using 25 months of LAT data, and found that 10 of the 13 sources decrease in significance and none of them increased substantially. Given the discussion above, the 13 unassociated source candidates with spectral indices in the range from 1.0 to 0.7 are consistent with being spurious and do not provide evidence for dark matter satellites.


%% file: figures.tex

\newcommand{\capThirteen}{Distribution of satellite mass and distance for the original VL-II satellites (in black) and the extrapolation to low-mass satellites (in red). Lower $J$-factors reside in the upper left while higher $J$-factors lie to the lower right. Contours of constant $J$-factor ($J\propto \frac{M^{0.81}}{D^2}$) run from the upper right to the lower left. One such contour is shown for the Draco dwarf spheroidal galaxy assuming a mass of $10^8\Msolar$ at a distance of $80\kpc$. Satellites lying in the hatched region above this line have lower $J$-factors than that of Draco.}
\ifcolor
   \ApJFigure{image13}{\capThirteen}{1.00}{image13}
\else
   \ApJFigure{image13_bw}{\capThirteen}{1.00}{image13}
\fi

\newcommand{\capOne}{Distribution of spectral indices and integral fluxes from $200\MeV$ to $300\GeV$ for the 385 high-latitude unassociated sources and source candidates. The squares are the 231 unassociated sources from the 1FGL catalog, while the triangles are the 154 additional source candidates detected with \Sourcelike.  The stars are the 10 representative power-law models in \Tabref{tests}.}
\ifcolor
  \ApJFigure{image1}{\capOne}{1.00}{image1}
\else
  \ApJFigure{image1_bw}{\capOne}{1.00}{image1}
\fi

\newcommand{\capSeven}{Best-fit exponentially cutoff power law (with $\Gamma {\rm =1.22}$ and $E_\mathrm{cut}=1.8\GeV$) of the millisecond pulsar 1FGL J0030+0451 (solid line) and the best-fit \bbbar spectrum (with $M_\mathrm{WIMP}=25\GeV$) of this pulsar (dashed line).}
\ApJFigure{image7}{\capSeven}{1.00}{image7}

\newcommand{\capEight}{Best-fit dark matter mass ($M_\mathrm{WIMP}$) coming from fitting 25 high-latitude ($|b|>20\degree$) pulsars with a \bbbar annihilation spectrum.}
\ApJFigure{image8}{\capEight}{1.00}{image8}


\newcommand{\capThree}{Distribution of spectral indices and integral fluxes from $200\MeV$ to $300\GeV$ for the 385 high-latitude unassociated sources and source candidates. The squares are the 231 unassociated sources from the 1FGL catalog, while the triangles are the 154 additional source candidates detected with \Sourcelike.  The circles are the 193 spurious sources found in a Monte Carlo simulation of background only.}
\ifcolor
  \ApJFigure{image3}{\capThree}{1.00}{image3}
\else
  \ApJFigure{image3}{\capThree}{1.00}{image3}
\fi


%% file: tables.tex

\begin{deluxetable}{ccccc}
\tablestyle{5}
\tablecaption{Values for \TSext{99} and \TSspec{99}}
\tablehead{
\colhead{Model Number} & \colhead{Spectral Index} & \colhead{Flux\tablenotemark{(a)}}             & \colhead{\TSext{99}} & \colhead{\TSspec{99}}  \\
                       &                          & \colhead{$(\photons \cm^{-2} \second^{-1})$} &                      & 
}
\startdata
1 & 0.9 & $2.0\times {10}^{-10}$ &  6.18  & 2.38 \\  
2 & 0.9 & $8.0\times {10}^{-11}$ &  7.87  & 2.46 \\  
3 & 1.5 & $1.1\times {10}^{-9}$  &  5.09  & 4.96 \\
4 & 1.5 & $2.0\times {10}^{-10}$ &  14.98 & 2.88  \\ 
5 & 2.0 & $1.2\times {10}^{-8}$  &  5.11  & 2.24 \\ 
6 & 2.0 & $1.2\times {10}^{-9}$  &  9.63  & 4.28 \\ 
7 & 2.5 & $2.1\times {10}^{-8}$  &  6.74  & 1.78 \\ 
8 & 2.5 & $0.5\times {10}^{-8}$  &  10.78 & 5.66  \\  
9 & 3.0 & $1.7\times {10}^{-8}$  &  9.81  & 2.14 \\ 
10& 3.0 & $1.0\times {10}^{-8}$  &  11.87 & 6.02  \\  
\enddata
\tablenotetext{(a)}{Integral flux from $200\MeV$ to $300\GeV$.}
\tablecomments{Cut values excluding point-like sources (\TSext{99}) and sources with power-law spectra (\TSspec{99}) at 99\% confidence. These values are independent of the $\TS > 24$ cut and were calculated from Monte Carlo simulations of the 10 typical power-law point source models.}
\label{tab:tests}
\end{deluxetable}


\begin{deluxetable}{lcccccc}
\tablestyle{7}
\tablecaption{Two Candidate Extended Sources}
\tablehead{
\colhead{Source ID} & \colhead{} & \colhead{ $l$, $b$\tablenotemark{(a)}} & \colhead{Flux\tablenotemark{(b)} }                    &\colhead{$\alpha_0$\tablenotemark{(c)}} & \colhead{$\TSext{}$}  & \colhead{$\TSspec{}$} \\
                    &            & \colhead{(deg)}                      & \colhead{($\photon \cm^{-2} \second^{-1}$)} &\colhead{(deg)}                     &                                          & }
\startdata
1FGL\,J1302.3$-$3255 & & 305.58, 29.90  & $1.33\times 10^{-8}$ & 1.2 & 9.3  & 4.6 \\    
1FGL\,J2325.8$-$4043 & & 349.83, $-$67.74 & $2.12\times 10^{-8}$ & 1.3 & 13.2 & $-$19.1 \\ 
\enddata
\tablenotetext{(a)}{Best-fit source position~\citep{1FGL}.}
\tablenotetext{(b)}{Integral flux from $200\MeV$ to $300\GeV$ interpolated from best-fit power-law model~\citep{1FGL}.}
\tablenotetext{(c)}{Best-fit spatial extension and $\pm 1 \sigma$ error using an NFW profile.}
\label{tab:extsources}
\end{deluxetable}


\begin{deluxetable}{lllccc}
\tablestyle{4}
\tablecaption{Detection Efficiency}
\tablehead{
& \colhead{Flux\tablenotemark{(a)}}                             &   &               & \colhead{Extension} &  \\
& \colhead{($\photons \cm^{-2} \second^{-1}$)}  &   & \colhead{0.5\degree} & \colhead{1.0\degree} & \colhead{2.0\degree} }
\startdata
& $0.2\times 10^{-8}$   &   & $<0.05$   &   $<0.05$   & $<0.05$     \\
& $0.5\times 10^{-8}$   &   & 0.16      &   0.28      & 0.31        \\    
& $1.0\times 10^{-8}$   &   & 0.74      &   0.76      & 0.83        \\
& $2.0\times 10^{-8}$   &   & 0.99      &   1.0       & 0.99        \\  
& $5.0\times 10^{-8}$   &   & 1.0       &   1.0       & 1.0         \\
\enddata
\tablenotetext{(a)}{Integral flux from $200\MeV$ to $300\GeV$.}
\tablecomments{Satellite detection efficiency for $100\GeV$ WIMP annihilating through the \bbbar channel.}
\label{tab:bbbar}
\end{deluxetable}
